\documentclass[aps,preprint,showpacs]{revtex4}%
\usepackage{amsfonts}
\usepackage{amsmath}
\usepackage{amssymb}
\usepackage{graphicx}%
\providecommand{\U}[1]{\protect\rule{.1in}{.1in}}

\begin{document}
\title{First-principles study of surface properties of PuO$_{2}$: Effects of
thickness and O-vacancy on surface stability and chemical activity}
\author{Bo Sun}
\affiliation{Institute of Applied Physics and Computational
Mathematics, Beijing 100094, P.R. China}
\author{Haifeng Liu}
\affiliation{Institute of Applied Physics and Computational
Mathematics, Beijing 100094, P.R. China}
\author{Haifeng Song}
\affiliation{Institute of Applied Physics and Computational
Mathematics, Beijing 100094, P.R. China}
\author{Guang-Cai Zhang}
\affiliation{Institute of Applied Physics and Computational
Mathematics, Beijing 100094, P.R. China}
\author{Hui Zheng}
\affiliation{Institute of Applied Physics and Computational
Mathematics, Beijing 100094, P.R. China}
\author{Xian-Geng Zhao}
\affiliation{Institute of Applied Physics and Computational
Mathematics, Beijing 100094, P.R. China}
\author{Ping Zhang}
\thanks{Corresponding author. Electronic mail: zhang\_ping@iapcm.ac.cn}
\affiliation{Institute of Applied Physics and Computational
Mathematics, Beijing 100094, P.R. China}


\begin{abstract}
The (111), (110), and (001) surfaces properties of PuO$_{2}$ are
studied by using density-functional theory+$U$ method. The
total-energy static calculations determine the relative order of
stability for low-index PuO$_{2}$ surfaces, namely, O-terminated
(111) $>$\ (110) $>$ defective (001) $>$ polar (001). The effect of
thickness is shown to modestly modulate the surface stability and
chemical activity of the (110) surface. The high work function
$\Phi$ of $6.19$ eV indicates the chemical inertia of the most
stable (111) surface, and the surface O-vacancy with concentration
$C_{\text{V}}$ = $25$\% can efficiently lower $\Phi$ to $4.35$ eV,
which is a crucial indicator of the difference in the surface
chemical activities between PuO$_{2}$ and $\alpha $-Pu$_{2}$O$_{3}$.
For the polar (001) surface, $50$\% on-surface O-vacancy can
effectively quench the dipole moment and stabilize the surface
structure, where the residual surface oxygen atoms are arranged in a
zigzag manner along the $<$100$>$ direction. We also investigate the
relative stability of PuO$_{2}$ surfaces in an oxygen environment.
Under oxygen-rich conditions, the stoichiometric O-terminated (111)
is found to be the most stable surface. Whereas under O-reducing\
conditions, the on-surface O-vacancy of $C_{\text{V}}$ = $1/9$ is
stable, and for high reducing conditions, the (111) surface with
nearly one monolayer subsurface oxygen removed ($C_{\text{V}}$ =
$8/9$) becomes most stable.

\end{abstract}
\maketitle

\section{Introduction}

Plutonium dioxide (PuO$_{2}$) is of high importance in the nuclear
fuel cycle and is particularly crucial in long-term storage of
Pu-based radioactive waste. Besides playing an important role in
both technological applications and environmental issues, Pu metal
and its oxides show many intricate physical behaviors due to the
complex electronic structure properties of 5f states
\cite{ref-1,ref-2,ref-3,ref-04}. Therefore, a thorough understanding
of the physical and chemical properties of PuO$_{2}$ is of great
significance, full of challenges and has always attracted attention.
Recently, there have occurred in the literature a series of
experimental reports \cite{ref-5,ref-6,ref-7} on the strategies of
storage of Pu-based waste. When exposed to air and moisture,
metallic plutonium surface rapidly oxidizes to PuO$_{2}$
\cite{Has2000,ref-4}. Under special aqueous condition, the
interaction of PuO$_{2}$ surface with adsorbed water can generate
non-stoichiometric PuO$_{2+x}$ (x$\leq$0.27) \cite{ref-5} via an
overall reaction, namely,
PuO$_{2}$+$x$H$_{2}$O$\rightarrow$PuO$_{2+x}$+H$_{2}$. However, the
oxidation of PuO$_{2}$ has been proved to be strongly endothermic by
subsequent first-principles theoretical calculation
\cite{PuO2+x,ref-8}. To test the possible existence of surface
PuO$_{2+x}$, recent photoemission study \cite{ref-7} has been
carried out and found that PuO$_{2}$ is only covered by a
chemisorbed layer of oxygen and can be easily desorbed at elevated
temperature. Thus, PuO$_{2}$ is generally acknowledged as the
highest stable Pu-oxide under ambient conditions. However, under
oxygen-lean conditions (in the vacuum or inert gas), the
PuO$_{2}$-layer can be \ reduced to sesquioxides (Pu$_{2}$O$_{3}$),
which can promote the corrosion of the Pu-metal by hydrogen
\cite{Has2000}. As we know, the low-temperature phase of
Pu-sesequioxide is a phase with space group $Ia\overline{3}$ (No.
206), which
is similar in the crystal structure to the cubic PuO$_{2}$\ ($Fm\overline{3}%
m$, No. 225) with the 25\% O vacancy located in the 16$c$
($0.25,0.25,0.25$) sites. The above mentioned experimental
observations indicate that the surface of PuO$_{2}$ is to some
extent chemically inactive, however, the formation of O vacancies
can prominently modify the electronic structure properties of\ both
the bulk and the surface of PuO$_{2}$. As a matter of fact, the
surface layers are directly involved in the significant corrosion
processes and many technological applications of the actinide
oxides, thus a deep understanding of the physical and chemical
properties of PuO$_{2}$ surfaces is always desirable. However, due
to the radioactivity and toxicity of Pu and the complexity of the Pu
element and Pu-O system, it is extraordinarily difficult to
experimentally explore the surface atomic and electronic structure
properties of the Pu-oxides, and particularly so for a single phase
compound.

From the theoretical point of view, conventional density-functional
theory (DFT) that applies the local density approximation (LDA) or
generalized gradient approximation (GGA) underestimates the strong
on-site Coulomb repulsion of the 5f electrons and, consequently,
describes PuO$_{2}$ as incorrect ferromagnetic FM conductor
\cite{ref-9} instead of antiferromagnetic AFM Mott insulator
reported by experiment \cite{ref-10}. Similar problems have been
confirmed in studying other correlated materials within the pure
LDA/GGA schemes. Fortunately, several approaches, including the
LDA/GGA+$U$ \cite{ref-11,ref-12,ref-13}, the hybrid density
functional of (Heyd, Scuseria, and Enzerhof) HSE \cite{ref-14}, the
self-interaction corrected local spin density SIC-LSD \cite{ref-15},
and LDA combined dynamical mean-field theory DMFT \cite{dmft}, have
been developed to correct these failures in calculations of actinide
compounds. The effective modification of pure DFT by LDA/GGA+$U$
formalisms has been confirmed widely in study of PuO$_{2}$
\cite{ref-16,ref-17,ref-18,ref-19,ref-20,ref-21}. By tuning the
effective Hubbard parameter in a reasonable range, the AFM Mott
insulator feature was correctly calculated and the structural
parameters as well as the electronic structure are well in accord
with experiments. However, those increasing theoretical researches
have been focusing on the bulk properties of PuO$_{2}$, and very
little is known regarding its surface physical and chemical
properties, which is in sharp contrast to the depth and
comprehensiveness of researches conducted upon the transition metal
and rare earth oxides \cite{SSR}. As far as we are aware, few DFT
studies of the PuO$_{2}$ (100) and (110) surfaces have been
presented in the literature \cite{ref-22,ref-23}. In addition, one
of the most important issues, i.e., the possible formation of O
vacancies and their effect on the atomic and electronic structures
of PuO$_{2}$ surfaces, remains completely unexplored.

Motivated by the above mentioned facts, in this paper, we
systematically study the surface properties of PuO$_{2}$.
Specifically, we have addressed (i) the structural stabilities of
the low-index PuO$_{2}$(111), (110), and (001) surfaces based on the
calculations of surface energies and surface relaxations, (ii) the
surface electronic properties such as the surface work function and
layer-projected density of the state (PDOS), and (iii) the effects
of Pu-oxide thickness and O-vacancy on the surface stabilities and
electronic structure properties.

This paper is organized as follows. The details of our calculations
are described in Sec. II. In Sec. III we present and discuss the
results. In Sec. IV, we summarize our main conclusions.

\section{Methodology of the calculation}

\subsection{DFT calculations}

The DFT calculations are carried out using the Vienna \textit{ab
initio} simulation package \cite{Vasp} with the frozen-core
projector-augmented wave (PAW) pseudopotentials \cite{Blo,Paw} and
plane-wave set. For the plane-wave
set, a cut-off energy of $400$ eV is used. The plutonium 6$s^{2}$7$s^{2}%
$6$p^{6}$6$d^{2}$5$f^{4}$, and the oxygen 2$s^{2}$2$p^{4}$ electrons
are treated as valence electrons. The exchange and correlation
effects are treated in both the LDA and the GGA \cite{Perdew}, based
on which the strong on-site Coulomb repulsion among the localized Pu
5f electrons is described by using
the LDA/GGA+$U$ formalisms formulated by Dudarev $et$ $al.$%
\cite{ref-11,ref-12,ref-13}. As concluded in some previous DFT
studies \cite{ref-16,ref-17,ref-19,ref-21}, although the pure LDA
and GGA fail to depict the electronic structure, especially the
insulating nature and the occupied-state character of bulk AFM
PuO$_{2}$, the LDA/GGA+$U$ approaches can capture the Mott
insulating properties of the strongly correlated Pu 5$f$ electrons
adequately and well reproduce experimental ground-state parameters
by tuning the effective Hubbard parameter $U_{\text{eff}}$ at
$\sim$4 eV. In this paper, the spherically averaged screened Coulomb
energy $U$ and the exchange energy $J$ for the Pu 5$f$ orbitals are
set to be 4.75 and 0.75 eV respectively, which have been tested and
applied in our previous studies of plutonium oxides
\cite{ref-16,ref-17,ref-21}. For fluorite structure PuO$_{2}$ of
AFM, our calculated equilibrium lattice parameter $a_{0}$=$5.466$
\AA \ within GGA+$U$ or $a_{0}$=$5.362$ \AA \ within LDA+$U$ is in
good agreement with the experimental value of $5.398$ \AA \
\cite{ref-5}. Our extensive test calculations in this work indicate
that the choice of $U_{\text{eff}}$ can alter the
electronic-structure properties of PuO$_{2}$ surfaces, as well as
those of bulk PuO$_{2}$. Specifically, when $U_{\text{eff}}$ is less
than 2.0 eV, the results of the electronic density of state (DOS)
indicate that PuO$_{2}$ surfaces are metallic FM-conductor instead
of the AFM Mott-insulators. The combination of $U$=4.75 and $J$=0.75
eV is also the optimum to well describe the electronic structure
properties of the surfaces of PuO$_{2}$, although the atomic
structural optimizations seem to be insensitive to the choice of
$U_{\text{eff}}$.

\begin{figure}[ptb]
\begin{center}
\includegraphics[width=0.7\linewidth]{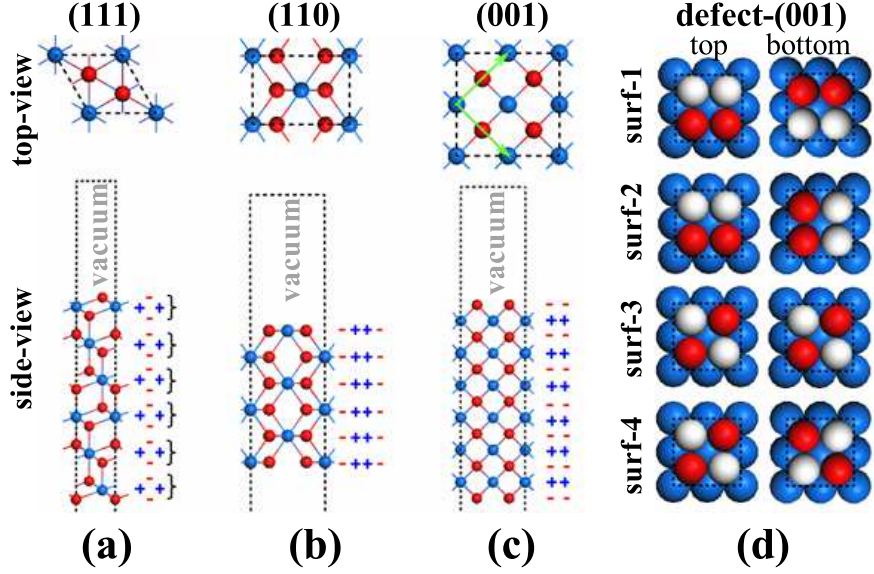}
\end{center}
\caption{(Color online) Low-index PuO$_{2}$ surface models: (a)
ideal
oxygen-terminated (111) unit cell; (b) (110) unit cell; (c) $(\sqrt{2}%
\times\sqrt{2})R45^{\text{o}}$ super-cell of ideal oxygen-terminated (001);
and (d) the defective (001) surface with four different distributions of 50\%
O-vacancy on the bottom and top faces, namely, surf-1, surf-2, surf-3 and
surf-4. The blue, red and white spheres denote Pu-atom, O-atom and O-vacancy,
respectively. In the side-views of three ideal surface, the labels of
different Pu-cation/O-anion $[++/-]$ stacking sequences are also indicated at
the lower part of the graphs. }%
\label{fig1}%
\end{figure}

The low-index surfaces of PuO$_{2}$ are modeled by finite-sized
periodic supercells, consisting of a number of oxide layers infinite
in $x$ and $y$ directions and separated in the $z$ direction by a
vacuum of 30 \AA . The Brillouin zone (BZ) integration is performed
using the Monkhorst-Pack (MP) $k$-point mesh \cite{Monk}.
Specifically, the two-dimensional (2D) unit $(1\times1)$ cell with a
$11\times11\times1$ $k$-point MP grid is employed for the
defect-free (ideal) PuO$_{2}$(111) and (110) surfaces, and a $(\sqrt
{2}\times\sqrt{2})R45%
{{}^\circ}%
$ supercell with a $7\times7\times1$ $k$-point grid is used for the
(001) surface. Examples of the initial slab configurations used in
the calculation are shown in Figs. 1(a)-(c), which are obtained by
truncating bulk PuO$_{2}$ along the [111], [110] and [001]
orientations, respectively. In the spin-polarized calculations with
the AFM order set to be in a simple
\textquotedblleft$\uparrow$ $\downarrow$ $\uparrow$ $\downarrow$%
\textquotedblright\ alternative manner along the $z$ direction, all
atoms are fully relaxed until the Hellmann-Feynman forces are less
than 0.01 eV/\AA . Various convergence tests have been performed to
ensure the above mentioned input parameters and models feasible and
reasonable in our current calculations. The result thereof shows
that the surface energy of a slab with certain thickness is
converged within 3 meV/\AA $^{2}$. In this paper, for the
convenience of depiction and plotting, we are using the number of
Pu-cation layer $N$ in a PuO$_{2}$\ slab to represent its thickness.
Thus, one can see in Figs. 1(a)-(c) that the ideal (111), (110), and
(001) slabs have the same thickness tag, namely, $N$= 6, and it is
worth noting that in practice these slabs consist of 18, 6 and 13
atomic monolayers (MLs), respectively.

According to Tasker's conclusion \cite{Tasker} on the surface
stabilities of ionic crystals, the side-view of three low-index
PuO$_{2}$ surfaces in Fig. 1 can reveal\ that the oxygen-terminated
(111) surface, consisting of successive and electrically neutral
\textquotedblleft O-Pu-O\textquotedblright blocks, is the
\textquotedblleft Tasker-type-II \textquotedblright\ surface, the
(110) is the typical \textquotedblleft type-I \textquotedblright\
surface stacked with identical neutral planes fulfilling the
PuO$_{2}$ stoichiometry, and the (001) is the typical
\textquotedblleft type-III \textquotedblright\ surface, consisting
of oppositely charged planes. Generally, type-I and type-II are
stable nonpolar surfaces, while type-III is unstable polar surface
due to the dipole moment of the repeated unit in $z$ direction. The
dipole moment may be quenched and the polar surface stabilized by a
variety of methods\cite{Polar-1,Polar-2,Polar-3,Polar-4}, including
surface reconstruction, the presence of adsorbates, and changes in
the surface electronic structure. In our present work, the polar
(001) surface is modeled as two oxygen-terminated surfaces with 50\%
oxygen vacancies to fulfill the stoichiometric formula, and the
different distributions of the 50\% O-vacancy considered here are
showed in Fig. 1(d). Besides above-mentioned three low-index
surfaces, the Pu-terminated (111) surface generated by removing the
outmost O layers of the O-terminated (111) surface has been found to
be quite unstable and eventually become O-terminated through a
significant reconstruction in the surface region due to the intense
dipole-dipole interaction. In such stable structure, the upper
layers resemble the $\beta$-Pu$_{2}$O$_{3}$(0001) surface. Thus, we
exclude the Pu-terminated (111) slab model and briefly name the
O-terminated PuO$_{2}$(111) surface as (111) surface if not
mentioned differently.

The surface energy $E_{\text{s}}$ is the energy needed to produce a
unit surface from a 3D infinite crystal and is one central quantity
in the studies of the relative stability of different surfaces. In
the DFT total-energy calculations of repeated slab-supercell
geometries, $E_{\text{s}}$ can be written as
\begin{equation}
E_{\text{s}}^{\text{relax/unrelax}}=\frac{1}{2A}\left(  E_{\text{slab}%
}^{\text{relax/unrelax}}-E_{\text{bulk}}\right)  , \label{e1}%
\end{equation}
where $E_{\text{slab}}^{\text{srelax/unrelax}}$ is the total energy
of the supercell with relaxed/unrelaxed slab, $E_{\text{bulk}}$ is
the energy of the reference bulk with the same number of atoms, and
the denominator $2A$ is the total area of both surfaces of the slab
with a finite thickness. Here, the convergence of the $E_{\text{s}}$
(i.e., $E_{\text{s}}^{\text{relax}}$ in Eq.(1)) is mainly determined
by two correlated factors, namely, the atomic structural relaxations
in several outmost layers and the thickness of the slab model. For
the surface structural relaxations, one can simply evaluate its
contribution to $E_{\text{s}}$ by calculating the surface relaxation
energy
$\Delta E_{\text{s}}=-(E_{\text{s}}^{\text{relax}}-E_{\text{s}}%
^{\text{unrelax}})$. The effect of the slab thickness should be
highlighted especially when the slab consists of the complicated
compounds such as the actinide oxides, and in addition, the work
functions calculated with slab approximations are known to be
depending on the slab thickness. Thus, in this work, both factors
will be considered and discussed in detail. \

\subsection{Thermodynamic considerations}

The DFT total-energy calculation gives $E_{\text{s}}$ only at zero
temperature $T$=$0$, zero pressure $P$=$0$, and for the surface in
contact with vacuum, which cannot be used to study the influence of
the realistic environmental conditions at a specific $T$ and $P$. To
further study the relative stability of the PuO$_{2}$ surfaces with
various concentrations of surface vacancy ($C_{\text{V}}$) at finite
$T$ and gas partial $P$ of the surrounding environment, we take the
approach of \textquotedblleft$ab$ $initio$ atomic
thermodynamics\textquotedblright\ \cite{abAT-1,abAT-2} to get the
surface Gibbs free energy $\gamma$($T$,$P$), with the general
expression given by
\begin{equation}
\gamma(T,P)=\dfrac{1}{2A}\left[  G(T,P,\{n_{\text{i}}\})-\sum_{\text{i}%
}n_{\text{i}}\mu_{\text{i}}(T,p_{\text{i}})\right]  , \label{e2}%
\end{equation}
where $G$ is the Gibbs free energy of the solid with the surface of
interest, $2A$ is the total surface area, $n_{\text{i}}$,
$\mu_{\text{i }}$and $p_{\text{i}}$\ are the particle number, the
chemical potentials and the partial pressures of the various
species. Here, the focus of our work is the relative stability of
O-terminated PuO$_{2}$ surfaces with different\ O-vacancy
concentrations, thus only two chemical species need to be
considered, namely i$=$Pu and O. In practice, the surface Gibbs
energy difference $\Delta\gamma(T,P)$ between a defective PuO$_{2}$
surface and the corresponding defect-free (ideal) surface is the
important quantity required, which can be written as
\begin{equation}
\Delta\gamma(T,P)=\dfrac{1}{2A}\left[  G^{\text{defect}}(T,P,N_{\text{V}%
})-G^{\text{ideal}}(T,P)+N_{\text{V}}\mu_{\text{O}}(T,p_{\text{O}_{\text{2}}%
})\right]  , \label{e3}%
\end{equation}
where $G^{\text{defect}}$ and $G^{\text{ideal}}$\ are the Gibbs free
energies of the supercells with the defective and ideal PuO$_{2}$
surfaces, respectively, and $N_{\text{V}}$ is the total number of O
vacancies on the PuO$_{2}$ surface. In the present situation, the
entropy and volume effects are small compared to the band energy in
the Gibbs free energy and thus are neglected in our calculations.
$\mu_{\text{O}}(T,p_{\text{O}_{\text{2}}})$ in
Eq. (3) is the oxygen chemical potential under partial pressure $p_{\text{O}%
_{\text{2}}}$ and for ideal oxygen-gase we can use the well-known
thermodynamic expression \cite{abAT-2}
\begin{equation}
\text{ }\mu_{\text{O}}(T,p_{\text{O}_{\text{2}}})=\frac{1}{2}\left(
E_{\text{O}_{2}}+\tilde{\mu}_{\text{O}_{2}}(T,p^{0})+k_{_{B}}T\ln
(p_{\text{O}_{\text{2}}}/p^{0})\right)  ,
\end{equation}
where $E_{\text{O}_{2}}$ is the total energy of the oxygen molecule.
For the
standard pressure $p^{0}=1$atm, the values of $\tilde{\mu}_{\text{O}%
_{\text{2}}}(T,p^{0})$ have been tabulated in Ref. 40. If we refer
the $\mu_{\text{O}}$\ to $\frac{1}{2}E_{\text{O}_{\text{2}}}$, then
the allowed
range for the $\Delta\mu_{\text{O}}=$\ $\mu_{\text{O}}-\frac{1}{2}%
E_{\text{O}_{\text{2}}}$ is given by \ %

\begin{equation}
-\frac{1}{2}E_{\text{f}}\leqslant\text{
}\Delta\mu_{\text{O}}\leqslant0,
\label{e8}%
\end{equation}
where $E_{\text{f}}$ is the formation energy of bulk PuO$_{2}$,
namely,
$E_{\text{f}}=\left\vert E_{\text{PuO}_{\text{2}}}-E_{\delta-\text{Pu}%
}-E_{\text{O}_{\text{2}}}\right\vert $.

To determine reasonable ranges of $\Delta\mu_{\text{O}}$, the
$\delta$-Pu is considered as reference system to calculate the
formation energy $E_{\text{f}}$ of bulk PuO$_{2}$. Since the
spin-orbit coupling (SOC) is important for certain properties of
heavy-metal compounds, we also include SOC effect in the
calculations of $E_{\text{PuO}_{\text{2}}}$ and $E_{\delta
-\text{Pu}}$. Finally, we restrict $\Delta\mu_{\text{O}}$ to $-4.89$
eV $\leqslant$ $\Delta\mu_{\text{O}}\leqslant0$\ based on the
GGA+$U$ and $-4.83$ eV $\leqslant$ $\Delta\mu_{\text{O}}\leqslant0$
based on the GGA+$U$+SOC.\ The
effect of spin polarization has been included in calculating $E_{\text{O}_{2}%
}$.

\section{Results and discussion}

\subsection{Surface energy and structural relaxation}

First, the relative stability of the low-index PuO$_{2}$ surfaces is
studied based on the systematic calculation of surface energy
$E_{\text{s}}$ and the detailed analysis of structural relaxation.
Furthermore, the effects of slab thickness on both surface stability
and relaxation are considered and discussed. In the following text,
we first present the results of non-polar (111) and (110) surfaces,
and then the polar (001) surface.

\begin{figure}[ptb]
\begin{center}
\includegraphics[width=0.7\linewidth]{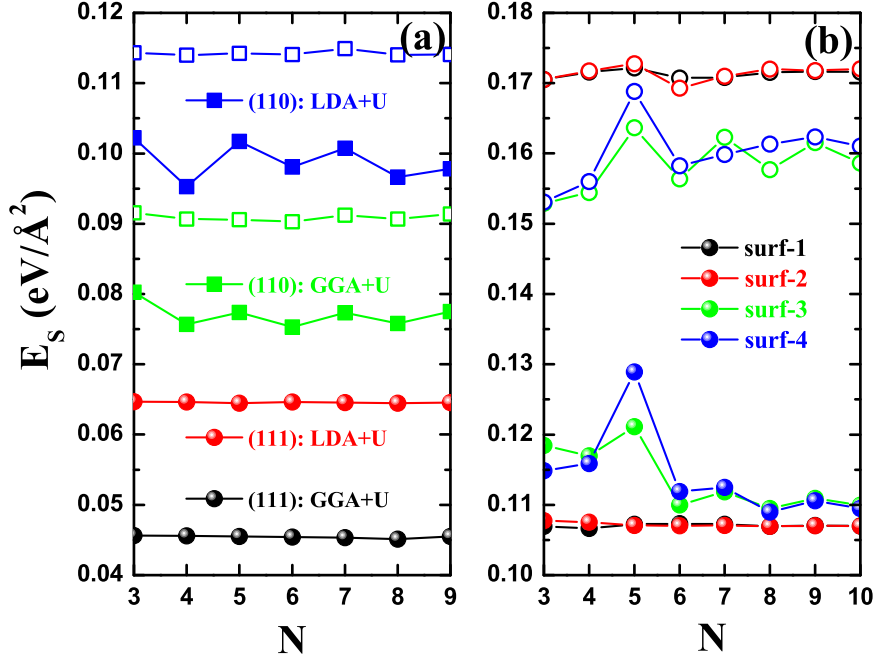}
\end{center}
\caption{(Color online) The surface energy $E_{\text{s}}$ as a function of the
PuO$_{2}$ slab thickness: (a) $E_{\text{s}}^{\text{relax}}$ (solid sphere) of
(111) surface, $E_{\text{s}}^{\text{relax}}$ (solid square) and $E_{\text{s}%
}^{\text{unrelax}}$ (open square) of the (110) surface; (b) $E_{\text{s}%
}^{\text{relax}}$ (solid sphere) and $E_{\text{s}}^{\text{unrelax}}$
(open circle) of defect-(001) surfaces. Note that the slab thickness
is defined by the number $N$ of the Pu-cation layers. Four different
terminations of
defect-(001) slabs in (b) are described in Fig. 1(d).}%
\label{fig2}%
\end{figure}

The calculated surface energy for fully relaxed\ (111) and (110)
slabs as a function of the thickness is illustrated in Fig. 2(a).
Obviously, both the GGA+$U$ and LDA+$U$ calculations give the
consistent results that the (110) surface energy is much higher than
the (111) surface energy. Generally, the calculated $E_{\text{s}}$
for (110) is 42\% with GGA+$U$ (or 33\% with LDA+$U$) higher than
for (111). Therefore, the PuO$_{2}$(111) surface is much more stable
than the \textquotedblleft more open\textquotedblright\ (110)
surface with relatively higher concentration of the surface dangling
bonds. For these two non-polar surfaces, Fig. 2(a) furthermore shows
two points worthy of special notice and further discussion: (i)
Despite the large difference in their respective $E_{\text{s}}$
between GGA+$U$ and LDA+$U$ calculations, the upper value (i.e., the
LDA+$U$ result) of the (111) surface is notably smaller than the
lower value (i.e., the GGA+$U$ result) of the (110) surface; (ii)\
The $E_{\text{s}}$ of the (111) surface is insensitive to the
thickness with steady values (0.045eV/\AA $^{2}$ for GGA+$U$ and
0.065eV/\AA $^{2}$ for LDA+$U$), whereas for the (110) surface the
evolution of $E_{\text{s}}$ as a function of slab thickness shows an
oscillating behavior, which indicates excellent agreement between
LDA+$U$ and GGA+$U$.

For the DFT energetic studies of solid materials, it is well known
that the GGA calculation usually underestimates the experimental
value and on the contrary the LDA often reports overestimated
results for many physical quantities, amongst which the surface
energy is a typical one. These opposite deviations from the
experimental measurement have been attributed to the
\textquotedblleft overbinding\textquotedblright\ character of LDA
and the consequent overcorrection of this defect in GGA
\cite{GC-1,GC-2}. However, we are more interested in comparing the
relative stabilities of different surfaces than in assessing the
different performances of LDA\ and GGA functionals, especially in
the absence of the experimental data. As far as we are aware, a few
existing DFT calculations have given a similar trend in LDA- and
GGA-$E_{\text{s}}$ results of metal oxides. For example, the LDA-$E_{\text{s}%
}$ results are 35\% and 22\% higher than the GGA results for
CeO$_{2}$(111) and (110) surfaces \cite{CeO2}, respectively, while
for MgO(001) surface the LDA-$E_{\text{s}}$ is 25.8\% higher than
the GGA-$E_{\text{s}}$ with the experimental values positioned in
between. Here, based on the above-mentioned point (i), we are
positive that the experimental observation will agree on the
prominent stability of the PuO$_{2}$(111) surface.

We now turn our attention to the evolution with the slab thickness
of the surface energies in Fig. 2(a). As is known, for a certain
cleaved surface, the consequent structural relaxation is an
effective way to minimize the surface cleavage energy, corresponding
to the $E_{\text{s}}^{\text{unrelax}}$, with
the contribution defined as the surface relaxation energy $\Delta E_{\text{s}%
}$=$-(E_{\text{s}}^{\text{relax}}\mathtt{-}E_{\text{s}}^{\text{unrelax}})$.
For (111) surface, the $E_{\text{s}}^{\text{unrelax}}$\ (not plotted
here) is very close to the $E_{\text{s}}^{\text{relax}}$, producing
a quite small $\Delta E_{\text{s}}$ less than 1.0 meV/\AA $^{2}$,
and the thickness effect on the surface relaxation can be neglected.
From Fig. 2(a), one can see that the $E_{\text{s}}^{\text{unrelax}}$
of (110) surface is clearly larger than the corresponding
$E_{\text{s}}^{\text{relax}}$ with their difference ($\Delta
E_{\text{s}}$) being higher than $10$ meV/\AA $^{2}$. Additionally,
we find that the $E_{\text{s}}^{\text{unrelax}}$ is to some extent
insensitive to the slab thickness, thus the oscillating behavior of
$E_{\text{s}}^{\text{relax}}$ is tied up to the dependence of
$\Delta E_{\text{s}}$ upon the slab thickness.

\begin{figure}[ptb]
\begin{center}
\includegraphics[width=0.75\linewidth]{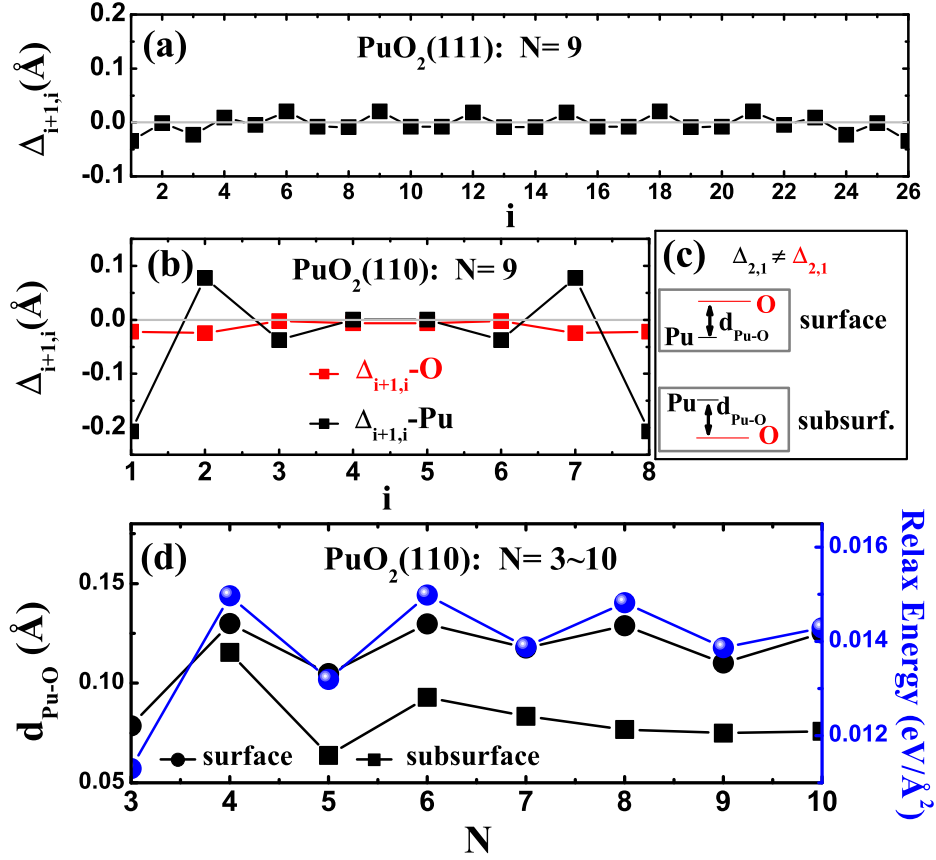}
\end{center}
\caption{ (Color online) The interlayer relaxation
$\Delta_{\text{i+1,i}}$ of (a) the (111) slab and (b) (110) slab
with $N=9$. (c) Sketch map of the vertical displacement
$d_{\text{Pu-O}}$ in the surface and subsurface of the relaxed (110)
slab due to the mismatch $\Delta_{\text{2,1}}$-Pu $\neq
\Delta_{\text{2,1}}$-O . (d) The $d_{\text{Pu-O}}$ in both surface
and subsurface of (110) slab and the relaxation surface energy
$\Delta
E_{\text{s}}$ as a function of the slab thickness.}%
\label{fig3}%
\end{figure}

In order to draw a clear comparison of the surface relaxation
between (111) and (110) surfaces\ and gain a detailed understanding
of \ the oscillating behavior of
(110)-$E_{\text{s}}^{\text{relax}}$, it proves to be quite necessary
to discuss the surface structures undergoing full relaxations.
Figures 3(a) and 3(b) show the interlayer relaxations along the
directions perpendicular to the (111) and (110) surfaces
respectively. Here the interlayer relaxation $\Delta_{\text{i+1,i}}$
is given by the optimized interlayer distance $d_{\text{i+1,i}}$\ in
a relaxed slab compared to the bulk interlayer distance
$d_{\text{i+1,i}}^{0}$ along the corresponding direction. Obviously,
the signs $+$ and $-$ of $\Delta_{\text{i+1,i}}$\ indicate expansion
and contraction of the interlayer spacing respectively. The stacking
sequence of the (111) slab with $N$=$9$ consists of 9
\textquotedblleft O-Pu-O\textquotedblright\ blocks, and therefore
this (111) slab contains totally 27 atomic monolayers (MLs). One can
see from Fig. 3(a) that the interlayer relaxations are really small,
so that the shrinkage ratio of the thickness is only
$\mathtt{\sim}$0.25\%. The (110) slabs used in Fig. 3(b) contain 9
atomic MLs with two oxygen and one plutonium atoms per 1$\times$1
cell being coplanar. Figure 3(b) shows that (i) the interlayer
relaxations in the (110) surface region of a few atomic layers are
prominently larger than those of the (111) surface in Fig. 3(a);
(ii) the interlayer relaxations of Pu sublattice are larger than the
oxygen sublattice, which is especially apparent in the outmost two
layers. Such mismatch of the relaxations between O- and
Pu-sublattices gives rise to the vertical displacement
$d_{\text{Pu-O}}$ between O and Pu atoms which are coplanar in the
unrelaxed (110) slab. Figure 3(c) gives a sketch map of the
$d_{\text{Pu-O}}$ in the surface and subsurface layers as a result
of the mentioned mismatch: $\Delta_{\text{2,1}}$-Pu
$\neq\Delta_{\text{2,1}}$-O. One can see that due to the nonzero
$d_{\text{Pu-O}}$, cation-anion dipoles with inverse orientations
are generated in the surface and subsurface respectively. Strictly
speaking, the relaxed (110) surface is now not a nonpolar surface
for Pu and O species. As we are aware, this observable surface
polarization of PuO$_{2}$(110) slab induced by the structural
relaxation is in good agreement with previous DFT calculation
\cite{ref-23}. Besides the interlayer (vertical) relaxation, the
inplane (lateral) relaxation of the surface layer (see Fig. 4(a))
tends to shorten the Pu-O bond on the surface by driving two
nearest-neighbor oxygen atoms (bonding to the same Pu atom) to close
up by $\mathtt{\sim}0.22$\r{A}, leading to the formation of O-O
dimers on the (110) surface. Furthermore, we have found that the
vertical displacement $d_{\text{Pu-O}}$\ shows an oscillating
behavior with the increasing slab thickness while the structure of
the inplane O-O dimers is to some extent insensitive to the slab
thickness. To reveal a clear-cut relationship between the structural
relaxation and the corresponding released energy $\Delta
E_{\text{s}}$, we plot the $d_{\text{Pu-O}}$ (in both surface and
subsurface) and $\Delta E_{\text{s}}$ as functions of the slab
thickness $N$ in Fig. 3(d). One can see that the oscillating
behaviors of $d_{\text{Pu-O}}$\ and $\Delta E_{\text{s}}$ are quite
in-phase, indicating that the oscillations of $E_{\text{s}}$ of
(110) surface as a function of slab thickness\ originate from the
interlayer relaxation.

\begin{figure}[ptb]
\begin{center}
\includegraphics[width=0.9\linewidth]{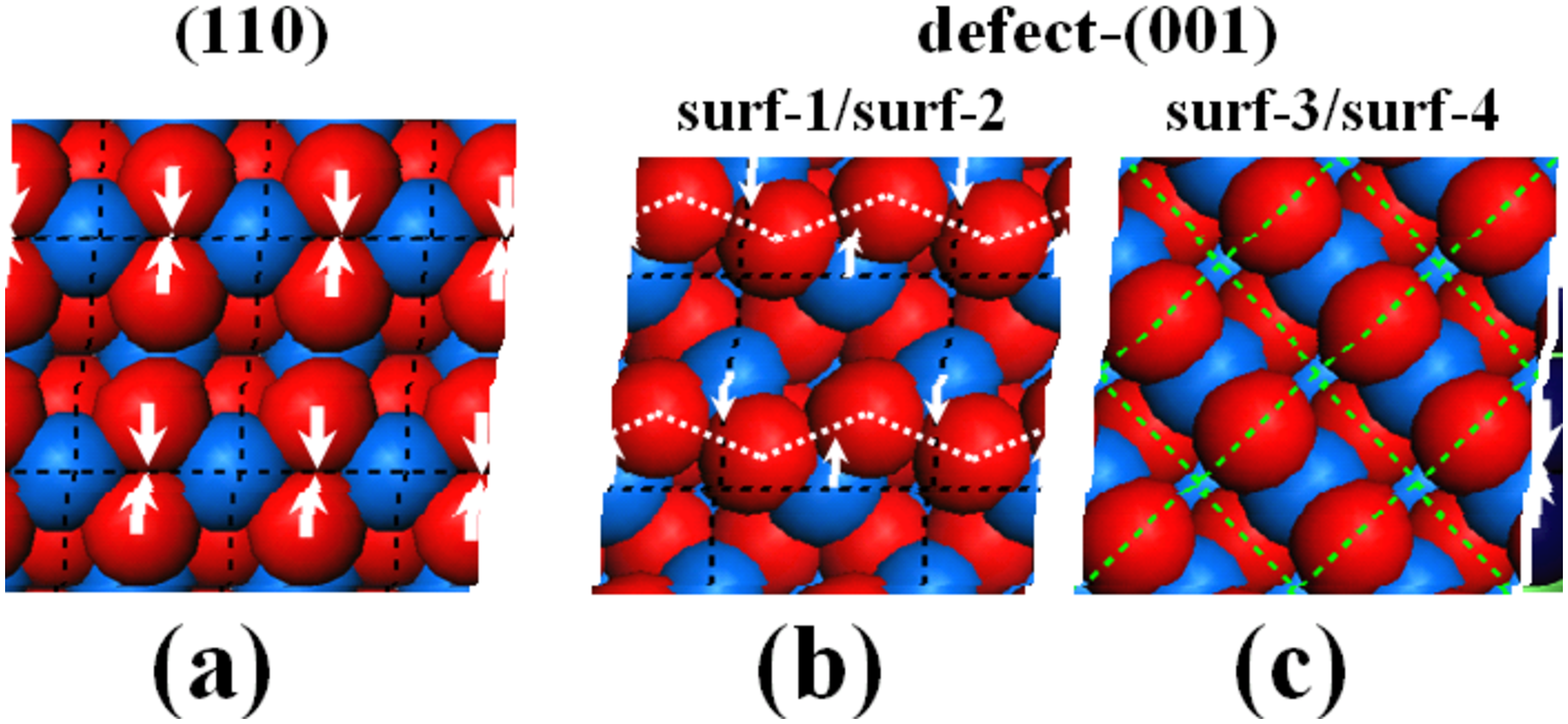}
\end{center}
\caption{ (Color online) The top view of the surface structures of
the relaxed (a) (110) slab, and the defect-(001) slabs: (b)
surf-1/surf-2 slabs and (c) surf-1/surf-2 slabs. The white arrows
(not to scale) indicate the directions of the inpane lateral
movements of the surface O atoms in (a) and
of the subsurface Pu-atoms in (b).}%
\label{fig4}%
\end{figure}

The ideal PuO$_{2}$(001) surface (see Fig. 1(c)) is an unstable
polar surface with an overall dipole field. However, it is found
that 50\% surface O vacancies in our defective (001) slab models
(see Fig. 1(d)) can effectively quench the dipole field and
stabilize the surface. Considering that a half oxygen vacancies can
usually induce the significant surface reconstruction, here we first
carry out the first-principles molecular dynamic (FPMD) simulations
based on GGA+$U$ within the micro-canonical ensemble to sufficiently
optimize the defect-(001) surface structures and then calculate
their zero-temperature surface energies. From the $E_{\text{s}}$
result (including $E_{\text{s}}^{\text{relax}}$ and
$E_{\text{s}}^{\text{unrelax}}$) presented in Fig. 2(b) as a
function of the slab thickness $N$, one can see that with increasing
$N$, the $E_{\text{s}}^{\text{relax}}$ for the four defect-(001)
surface models converges to $\mathtt{\sim}0.11$eV/\AA $^{2}$, which
is still notably larger than that of the (110) surface.
Interestingly, according to the results of
$E_{\text{s}}^{\text{relax}}$, surf-3 and surf-4 slab models are a
bit less stable than surf-1 and surf-2 in the whole range of
slab thickness that we considered. However, the values of $E_{\text{s}%
}^{\text{unrelax}}$ for surf-3 and surf-4 models are lower than
those for surf-1 and surf-2 models mainly due to the two different
distributions of the surface oxygen vacancies, namely, the
missing-row and uniform types in surf-1/surf-2 and surf-3/surf-4
respectively.

After the structural optimization by the FPMD simulations, for all
four defect-(001) surface models with $N=8$, the surface oxygens as
well as the subsurface oxygens beneath relax inward by $0.26$ \AA \
and $0.14$ \AA \ respectively, while the subsurface oxygens without
surface oxygen above relax outward by $\mathtt{\sim}0.17$ \AA. For
surf-1/surf-2, the Pu-sublattice relaxes inward by
$\mathtt{\sim}0.02$ \AA, on the contrary, the Pu-sublattice relaxes
outwards by $0.05$ \AA \ for surf-3/surf-4. Accompanied with a
slight discrepancy in vertical relaxations of the Pu-sublattice for
surf-1/surf-2 and surf-3/surf-4, it is found that the difference in
$\Delta E_{\text{s}}$\ for these two types of (001) terminations is
mainly caused by the distinguishing inplane (lateral) relaxations of
subsurface Pu-sublattices, which are sketched in Figs. 4(b) and
4(c). For surf-1/surf-2, the Pu atoms bonding to the same surface
oxygen atom relax to close up by $\mathtt{\sim}0.45$ \AA \ and this
periodic lattice distortion consequently provoke the zigzag manner
reconstruction of surface oxygen-lattice from the initial linear
chain. Interestingly, this peculiar reconstruction was experimentaly
observed in the defective polar (001) surface of uranium dioxide
UO$_{2}$ with 50\% oxygen vacancies \cite{UO2-zigzag}. However,
because of the uniform distribution of the surface O-vacancies, the
Pu and O atoms in surf-3/surf-4 keep lateral inaction. For the
defective (001) surfaces, our current results show that both the
slab thickness and the distributions of the surface O-vacancies can
notably impact the surface stability, and there may be several more
stabilizing mechanisms coexisting on the polar (001) surface. \

To briefly summarize our results in this section, we give the
relative order of stability for low-index PuO$_{2}$ surfaces,
namely, (111) $>$\ (110) $>$ defect-(001) $>$ polar-(001), which is
well consistent with that of CeO$_{2}$ \cite{abAT-6,CeO2} and
UO$_{2}$ \cite{UO2}, which are of the same fluorite structure as
PuO$_{2}$.

\subsection{Surface electronic structure and work function}

The bulk PuO$_{2}$ is considered to be the AFM Mott insulator
according to the experimental report \cite{ref-10}. Here the
atom-projected density of the electronic states (PDOSs) for the Pu
and oxygen atoms in the bulk PuO$_{2}$ and on the relaxed (111),
(110), and defect-(001) surfaces are shown in Fig. 5. Since the
GGA+$U$ and LDA+$U$ give the similar description of the PDOS, here
we only plot the GGA+$U$ results. The orbital-resolved PDOS of the
bulk PuO$_{2}$ at the ground state has been calculated and analyzed
in detail by previous DFT+$U$ \cite{ref-16,ref-18,ref-19,ref-21} and
hybrid DFT \cite{ref-4,ref-14} studies, and those theoretical
results of DOS are usually tested by comparing with the experimental
photoelectron spectroscopy (PES) measurements \cite{ref-6,ref-7}.

\begin{figure}[ptb]
\begin{center}
\includegraphics[width=0.7\linewidth]{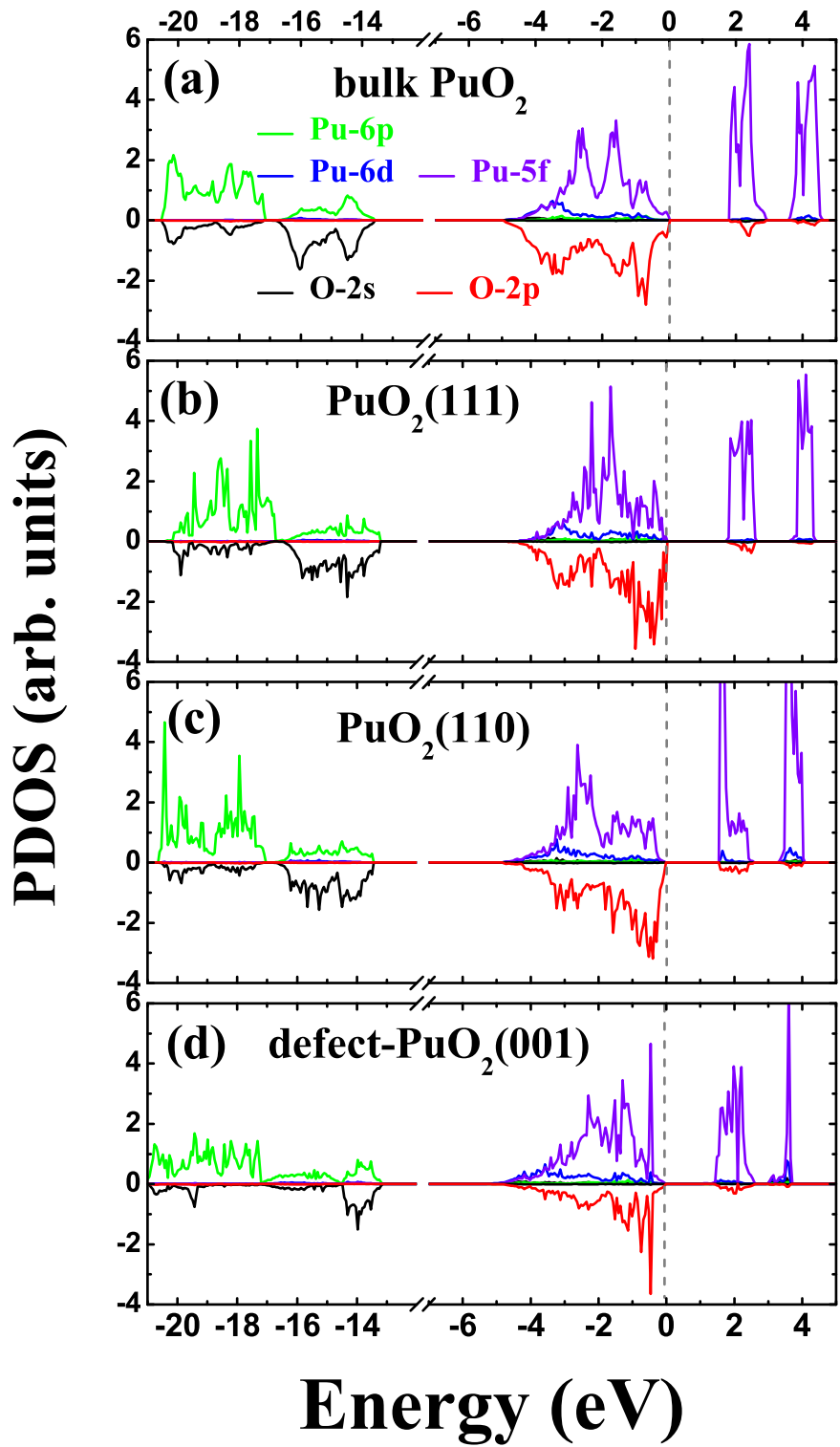}
\end{center}
\caption{ (Color online) The atom-projected (orbital resolved) DOS
for (a) bulk PuO$_{2}$, (b) (111) surface, (c) (110) surface, and
(d) defect-(001)
surface. The Fermi level is indicated by the vertical dashed line at $0$ eV.}%
\label{fig5}%
\end{figure}

Here, we replot the PDOS of the bulk PuO$_{2}$ with AFM\ phase as a
benchmark for those of the Pu-O atoms on different PuO$_{2}$
surfaces, aiming at finding significant influence in the electronic
structure by the inclusion of surface effect. The PDOS of bulk
PuO$_{2}$ in Fig. 5(a) demonstrates the following features: (i)
Above the Fermi level, the insulating band gap is about $1.8$ eV,
which is in good agreement with the experimental measurement
\cite{ref-6}; (ii) Below the Fermi level the highest occupied band
(HOB) with a range of $-5$ to $0$ eV is mainly the $5f$(Pu)-$2p$(O)
hybridization, with little contributions from $6p$ and $6d$ states
of Pu; (iii) The lower occupied band (LOB) with a range of $-21$ to
$-13$ eV consists of Pu-$6p$ and O-$2s$ states. \

From the surface DOS in Fig. 5, one can see that the PDOS
distribution for the (111) surface shows a close resemblance to the
case of the bulk PuO$_{2}$, specifically, the similarities in the
insulating band gap and the structures of Pu-$5f$ state with two
pronounced peaks are so strong that the slight contraction and
shift-up of the HOB are almost covered up. For the (110) surface,
the insulating band gap reduces to $\mathtt{\sim}1.6$ eV, and
particularly, the two-peak structure of Pu-$5f$ disappears mainly
due to the existing surface polarization with nonzero
$d_{\text{Pu-O}}$, which modifies the crystal symmetry of the oxide
surface layer. For the (001) surface, since the surface layer of
surf-1 slab model used in Fig. 5(d) is not the stoichiometric
PuO$_{2}$ but the `Pu-O', the insulating band gap further reduces to
$\mathtt{\sim}1.4$eV, and a sharp peak of Pu-$5f$ state emerges
below the Fermi level, which implies the increase in the localized
correlation of the Pu-$5f$ electronic state due to the presence of
oxygen vacancies. These facts suggest that the surface effect of
PuO$_{2}$ upon the electronic structure of the bulk phase appears to
be insignificant for the stable (111) surface, to some extent
significant for the (110) surface, and remarkable for the
defect-(001) surface. In order to be able to theoretically reproduce
the whole PES spectra of PuO$_{2}$ layers by the right description
of the complex behavior of Pu-$5f$ state, there is clearly much left
to be done.

\begin{figure}[ptb]
\begin{center}
\includegraphics[width=0.7\linewidth]{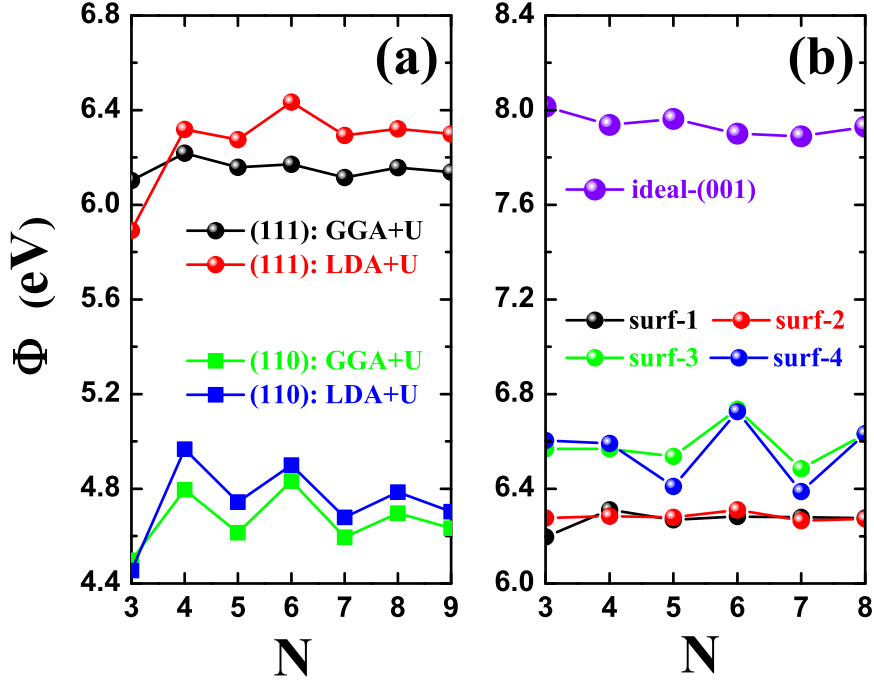}
\end{center}
\caption{ (Color online) The calculated work function $\Phi$ as a
function of the slab thickness: (a) PuO$_{2}$(111) and (110)
surfaces, (b) the ideal (001)
and the four defect-(001) surfaces.}%
\label{fig6}%
\end{figure}

In addition to the PDOS, we have also calculated the work function
$\Phi$ of low-index PuO$_{2}$ surfaces, and plotted them in Fig. 6.
The work function $\Phi$ is the minimum energy required to remove an
electron from the surface to the vacuum and can be written as
$\Phi=V_{\text{vacuum}}-$\ $E_{\text{F}}$, where $V_{\text{vacuum}}$
is the planar-averaged electrostatic potential in the middle of the
vacuum and $E_{\text{F}}$ is the Fermi energy of the system.
Therefore, work function is one fundamental physical quantity for
the surface reactivity. Furthermore, a modified or tunable $\Phi$
can be useful for applications such as catalysis, because a slight
change in the energy scale is exponentially amplified for chemical
reactions.

From Fig. 6(a), one can see that the $\Phi$ of stable (111) surface
with an average value of $\mathtt{\sim}6.1$ eV (GGA+$U$) or
$\mathtt{\sim}6.3$ eV (LDA+$U$) is much higher than that of the
(110) surface with an average value of $\mathtt{\sim}4.7$ eV
(GGA+$U$) or $\mathtt{\sim}4.8$ eV (LDA+$U$). Thus, for nonpolar
surfaces, stable (111) surface will show its inertness in the
surface chemical reactions, and the more open (110) surface is
expected to be chemically active.\ Figure 6(a) also demonstrates the
convergence behavior of $\Phi$ as a function of the slab thickness.
Here, it is found that the GGA+$U$ and LDA+$U$ results of $\Phi$ are
in general agreement. For the stable (111) surface, the work
function shows less responsiveness to the thickness effect, as well
as its surface energy $E_{\text{s}}$ in Fig. 2(a). On the contrary,
both $\Phi$ and $E_{\text{s}}$ of the (110) surface are sensitive to
the slab thickness with a convergent oscillation. Combining with the
surface relaxation results given in Fig. 3(d), one can conclude that
the thickness effect modifies both the surface stability and surface
chemical activity through the structural relaxations. Usually, the
thickness of the oxide film formed on Pu metal during stockpile
process is typically of nanometer scale. Thus, our present finding
of thickness-selective surface activity may help to deepen the
understanding of the microscopic mechanisms for the chemical
reaction of small molecules (such as water) on oxidized Pu surfaces,
which is fundamental to the safety issue of nuclear industry.
Interestingly, a recent DFT study \cite{MgO} has reported that the
thickness effect of MgO film can be used to control the dissociation
of water molecule on surface.

For the case of the polar (001) surface shown in Fig. 5(b), the work
function is mainly dominated by the strength of the anion-cation
dipole, which impedes the escape of electrons. Therefore, one can
see that the work function of the ideal (001) surface is close to
$8$ eV, which is much higher than that of the defect-(001) surfaces
with an average value of $\mathtt{\sim}6.4$ eV. Due to the
significant influence of the existing dipole on the (001) surface,
it is not reasonable to compare its surface chemical activity with
ideal (110) or (111) non-polar surface merely via the surface work
function.

\subsection{Effect of oxygen vacancy}

In this section, we focus on the effect of O-vacancy with various
concentrations upon the surface activity and surface relative
stability by using the static GGA+$U$ calculation and the approach
of \textquotedblleft$ab$ $initio$ atomistic
thermodynamics\textquotedblright. Here, our current study is mainly
driven by the following motivations: (i) to explain the difference
in the surface chemical activity between PuO$_{2}$ and $\alpha$-Pu$_{2}$%
O$_{3}$, as mentioned in Sec-I; (ii) to discuss the mechanism of
creating surface oxygen-vacancy in the cancelation of the polarity
for PuO$_{2}$(001) surface; and (iii) to explore the stable surface
phase of PuO$_{2}$ in an oxidizing environment. Amongst these listed
issues, the O-vacancy is obviously the major factor.

In the calculation, to eliminate the thickness effect of the
PuO$_{2}$ slab, we employ (111), (110), and (001) slabs with $N=6$,
$8$, and $8$ respectively. The various concentrations of oxygen
vacancy ($C_{\text{V}}$) are modeled by removing different amounts
of oxygen atoms from ideal slabs with a series of surface unit
cells. Here, $C_{\text{V}}$ is the ratio between the number of
O-vacancies and the total number of O atoms on the ideal surface
layer. Specifically, for the (111) surface, slabs of $(2\times2)$
and $(3\times3)$ unit cells are employed to create $C_{\text{V}}$ of
$\frac{1}{9}$, $\frac {1}{4}$, $\frac{1}{2}$, $\frac{3}{4}$,
$\frac{8}{9}$, and $1.0$. For the (110) surface, slab of
$(1\times2)$ unit cells is used to create $C_{\text{V}}$ of
$\frac{1}{4}$, $\frac{1}{2}$, $\frac{3}{4}$, and $1.0$. For the
(001) surface,
slab of $(\sqrt{2}\times\sqrt{2})R45%
{{}^\circ}%
$ unit cells can create $C_{\text{V}}$ of $\frac{1}{4}$,
$\frac{1}{2}$, $\frac{3}{4}$, and $1.0$.

Before the discussion on the evolution of work function as a
function of $C_{\text{V}}$, we first present the O-vacancy effect
upon the structural relaxation of the (111) slab. The static
calculation demonstrates that the (111) surface with on-surface
O-vacancy is slightly preferred in total energy
when $C_{\text{V}}\leqslant$1/2. However, when 1/2%
$<$%
$C_{\text{V}}\leqslant$1, the subsurface oxygen atoms, each of which
sharing the same unit surface cell with one certain surface oxygen,
will break through the above Pu-terminated layer to form a complete
O-terminated surface, leaving the subsurface O-vacancies at their
former sites. Therefore, when
$C_{\text{V}}$%
$>$%
1/2, all the on-surface O-vacancies will convert to be the
subsurface O-vacancies by a significant reconstruction.

\begin{table}[ptb]
\caption{The calculated work function $\Phi$ (in eV) of PuO$_{2}$(111), (110)
and (001) surfaces. }%
\label{table1}
\begin{tabular}
[c]{ccccccc}\hline\hline
& \ $C_{\text{V}}$=0 \  & \ $C_{\text{V}}$=1/9 \  & \ $C_{\text{V}}$=1/4 \  &
\ $C_{\text{V}}$=1/2 \  & \ $C_{\text{V}}$=3/4 \  & \ $C_{\text{V}}$=1
\ \\\hline
(111): on-surface & 6.19 & 5.09 & 4.35 & 4.07 & --- & ---\\
(111): subsurface & 6.19 & 5.36 & 5.18 & 4.87 & 4.49 & 2.57\\
(110): on-surface & 4.70 & --- & 3.84 & 3.57 & 2.80 & 2.44\\
(001): on-surface & 7.93 & --- & 7.24 & 6.63 & 4.68 & 3.25\\\hline\hline
\end{tabular}
\end{table}

The GGA+$U$ calculated work function $\Phi$\ of (111), (110), and
(001) surfaces for different values of $C_{\text{V}}$ is listed in
Table I, where the \textquotedblleft on-surface\textquotedblright\
and the \textquotedblleft subsurface\textquotedblright\ denote the
initial pure on-surface and subsurface distributions of O-vacancies
for the (111) surface. One can see from Table I that for all three
surfaces the work function will monotonically reduce with increasing
Cv. Therefore, the introduction of O-vacancy can prominently enhance
the surface chemical activity of non-polar (111) and (110) surfaces.
For instance, the work function of the ideal (111) surface is $6.19$
eV, while a low $C_{\text{V}}$ = $1/9$ $(1/4)$ of on-surface
O-vacancy can effectively depress the work function by $1.1$
$(1.84)$ eV and efficiently amplify the probability of the chemical
reaction between the PuO$_{2}$(111) surface and other small gaseous
molecules such as H$_{2}$ and H$_{2}$O, which will be investigated
in our next work. This result can be also extended to explain the
significant difference in the chemical activities between PuO$_{2}$
and $\alpha$-Pu$_{2}$O$_{3}$, the latter has been found in
experiment \cite{Has2000} to be more active in interacting with
small molecules.

\begin{figure}[ptb]
\begin{center}
\includegraphics[width=0.9\linewidth]{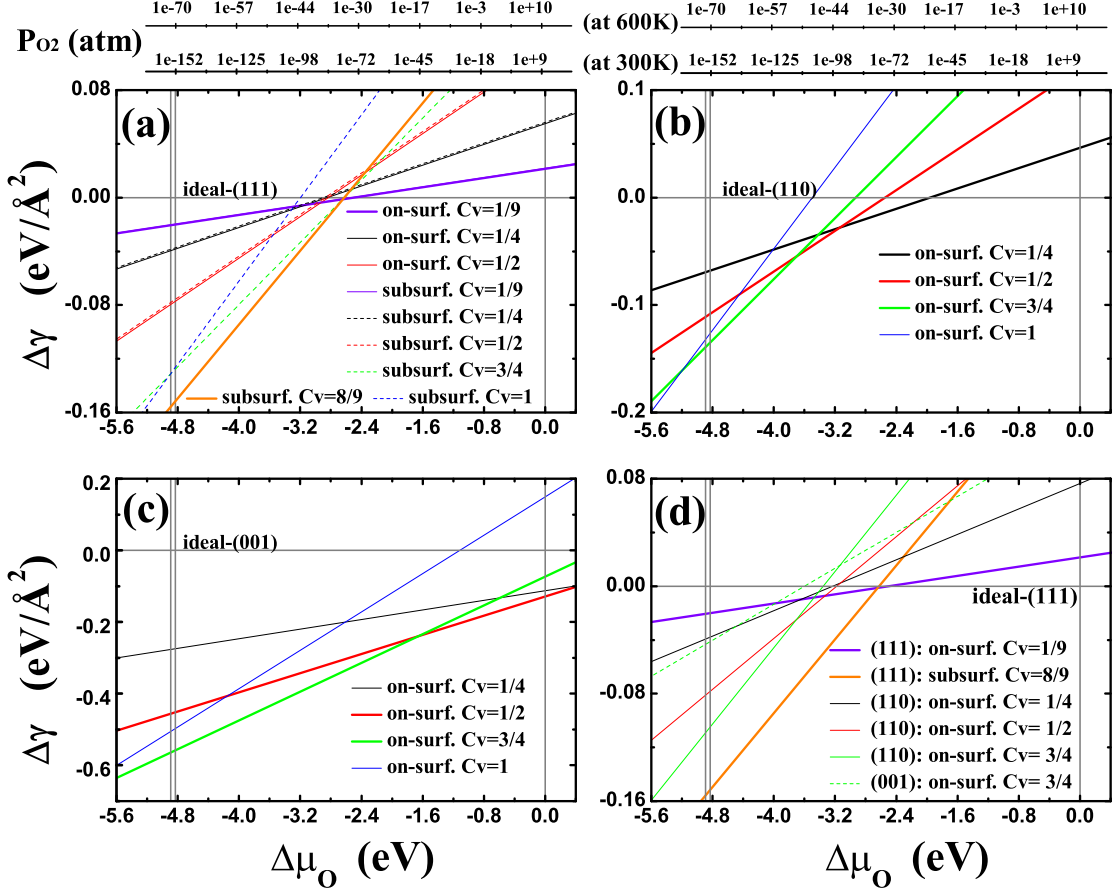}
\end{center}
\caption{(Color online) Surface free energy difference
$\Delta\gamma$ of (a) PuO$_{2}$(111), (b) (110) and (c) (001)
surfaces with various concentrations of O-vacancy Cv as a function
of the oxygen chemical potential $\Delta \mu_{\text{O}}$, with the
corresponding pressure lines at T=300 K and T=600 K. The low-energy
surface terminations are drawn with the thick lines in (a)-(c),
which are gathered in (d) with the ideal (111) surface as the
reference
configuration. }%
\label{fig7}%
\end{figure}

Assuming that the PuO$_{2}$ surface is in equilibrium with an
external oxygen environment and translating the oxygen chemical
potential into temperature and pressure conditions according to Eq.
(3) and Eq. (4) respectively, we first discuss effect of the
O-vacancy upon the relative stability of one certain surface.
Figures 7(a)-(c) present the Gibbs surface free energy difference
$\Delta\gamma$ of (111), (110) and (001) surfaces respectively. Here
the $\Delta\gamma$ is calculated by Eq. (3) with the corresponding
ideal surface as the reference system. For the (111) surface, from
Fig. 7(a) one can see that (i) the ideal, vacancy-free (111) surface
is the most stable structure under the oxygen rich conditions with
$\Delta\mu_{\text{O}}\geqslant-2.49$ eV; (ii) then defect-(111)
surface with on-surface O-vacancy of $C_{\text{V}}$ $=1/9$ becomes
stable within a very narrow range of $-2.63$ $\leqslant$
$\Delta\mu_{\text{O}}<-2.49$ eV; (iii) for further reducing
environment with $\Delta\mu_{\text{O}}\leqslant$ $-2.63$ eV, the
defect-(111) surface with subsurface O-vacancy of high
$C_{\text{V}}$=$8/9$ becomes the most stable.

For the (110) surface, Fig. 7(b) shows that (i) the ideal (110)
surface is most stable within the range of the
$\Delta\mu_{\text{O}}\geqslant-1.96$ eV; (ii) the defect surface
with $C_{\text{V}}$=$1/4$ becomes most stable within $-2.54$
$\leqslant$ $\Delta\mu_{\text{O}}<-1.96$ eV; (iii) the defect
surface with $C_{\text{V}}$=$1/2$ in succession becomes most stable
when $-2.93$ $\leqslant$ $\Delta\mu_{\text{O}}<-2.54$ eV; (iv) the
defect surface with $C_{\text{V}}$= $3/4$ becomes most stable when
$-3.50$ $\leqslant$ $\Delta \mu_{\text{O}}<-2.93$ eV, and the
Pu-terminated (110) surface with $C_{\text{V}}$= $1$ is not a stable
surface phase in the whole range of the allowed $\Delta
\mu_{\text{O}}$.

From the results of polar-(001) surface in Fig. 7(c), we have found
that (i) under the O-rich conditions, the ideal (001) surface is
unstable, however the $50\%$ surface O-vacancies can efficiently
stabilize the polar surface (ii) when $\Delta\mu_{\text{O}}$
$\leqslant-1.69$ eV, the defect-(001) surface with
$C_{\text{V}}$=$3/4$ is the optimal surface structure; (iii) within
the whole allowed range of $\Delta\mu_{\text{O}}$, the Pu-terminated
(001) surface is unstable so as the vacancy-free O-terminated (001)
surface.

We collect all the stable surface phases from Fig. 7(a)-(c), and
summarize them in Fig. 7(d) by taking the ideal (111) surface as the
reference structure. One can see that the ideal (111) surface,
defect-(111) surface with low on-surface O-vacancy concentration of
$C_{\text{V}}$= $1/9$ and with high sub-surface $C_{\text{V}}$
=$8/9$ are the stable surface structures. That is to say, the ideal
(111) surface is stable under the oxygen-rich conditions, while for
an oxygen-reducing environment the (111) surface with nearly one
monolayer subsurface oxygen removed become stable, and the
on-surface oxygen vacancy with low $C_{\text{V}}$ of $1/9$ can
minimize the Gibbs surface energy in a very narrow range of
$\Delta\mu_{\text{O}}$.

\begin{table}[ptb]
\caption{The calculated formation energies of the O-vacancy $E_{\text{v}}$ (in
eV).}%
\label{table2}
\begin{tabular}
[c]{ccccccc}\hline\hline
& \ $C_{\text{V}}$=1/9 \  & \ $C_{\text{V}}$=1/4 \  & \ $C_{\text{V}}$=1/2
\  & \ $C_{\text{V}}$=3/4 \  & \ $C_{\text{V}}$=8/9 \  & \ $C_{\text{V}}$=1
\ \\\hline
(111): on-surface & 2.49 & 2.85 & 2.83 & --- & --- & ---\\
(111): subsurface & 2.54 & 2.89 & 2.87 & 2.63 & 2.63 & 3.20\\
(110): on-surface & --- & 1.96 & 2.54 & 2.93 & --- & 3.50\\
(001): on-surface & --- & -3.38 & -1.93 & -0.73 & --- & 1.11\\\hline\hline
\end{tabular}
\end{table}

In Table II, we have listed the O-vacancy formation energies
$E_{\text{V}}$, which can be defined as
$E_{\text{v}}=\frac{1}{N_{\text{O-V}}}\left[
E_{\text{slab}}^{\text{defect}}-E_{\text{slab}}^{\text{ideal}}+N_{\text{O-V}%
}\cdot\frac{1}{2}E_{\text{O}_{\text{2}}}\right]  $, where
$N_{\text{O-V}}$ is
the total number of the O-vacancy in a defective slab, $E_{\text{slab}%
}^{\text{defect}}$, $E_{\text{slab}}^{\text{ideal}}$ and $E_{\text{O}%
_{\text{2}}}$ are the total energies of the defective slab, ideal
slab and a free oxygen molecule, respectively. One can see for the
(111) surface that the $E_{\text{V}}$ does not show considerable
change except in the extreme case of $C_{\text{V}}$ = 1 with a
maximum value $E_{\text{V}}$= $3.20$ eV. On the contrary, the
$E_{\text{V}}$ of the (110) surface is sensitive to the
$C_{\text{V}}$, namely, the $E_{\text{V}}$ monotonically increases
with increasing $C_{\text{V}}$, indicating a notable interaction
between the vacancies. Finally, the polar (001) surface is a special
case. The minus $E_{\text{V}}$ indicates that the formation of
surface O-vacancy is an exothermic process, and at the same time
stabilizes the polar surface. However the $E_{\text{V}}$ also
monotonically increases with the $C_{\text{V}}$ and rises to $+1.11$
eV when $C_{\text{V}}=1$. \

\section{Conclusions}

To conclude, we have systematically studied the basic surface
properties of low-index PuO$_{2}$(111), (110), and (001) surface by
means of the first-principles DFT calculations within the LDA+$U$
and GGA+$U$ frameworks. The defect-free O-terminated (111) surface
is found to be most stable, possessing the lowest E$_{s}$ that is
insensitive to the thickness of the film. The surface energy of the
non-polar (110) surface is 33\% to 42\% higher than that of the
(111) surface, accompanying with an oscillating behavior with the
film thickness. The polar (001) surface has been modeled using 50\%
oxygen vacancies to cancel the polarity. The residual surface oxygen
atoms have been found to reconstruct in a zigzag manner along the
$<$%
100%
$>$
direction. In connected with the relative order of stability for
these three low-index surfaces, our calculated surface electronic
structures have displayed from insignificant to remarkable deviation
from the bulk case. The work function $\Phi$ has also been
systematically investigated, and a high value of about $6.19$ eV for
the most stable (111) surface indicates its low chemical activity.
Remarkably, this value can be reduced to $4.35$ eV with 25\%
oxygen-vacancy present on the surface. This conclusion can be used
to explain the difference in the surface chemical activities between
PuO$_{2}$ and $\alpha$-Pu$_{2}$O$_{3}$.

We have also investigated the surface thermodynamics in an oxygen
environment. Our results have indicated that under oxygen-rich
conditions, the stoichiometric (111) surface is most stable. Under
oxygen-reducing conditions, the on-surface O-vacancy of low
concentration $C_{\text{V}}$ = $1/9$ can slightly minimize the Gibbs
surface energy $\gamma$ of (111) in a narrow range of the oxygen
chemical potential $\Delta\mu_{\text{O}}$. For the highly reducing
conditions, the (111) surface with nearly one monolayer subsurface
oxygen removed ($C_{\text{V}}$ = $8/9$) becomes most stable, where
the upper layers resemble the $\beta$-Pu$_{2}$O$_{3}$(0001) surface.
Based on these systematic results, our current study may provide a
guiding line to understand various chemical properties and processes
occurred on PuO$_{2}$ surfaces.

\begin{acknowledgments}
This work was supported by NSFC under Grant No. 51071032, and by the
Foundations for Development of Science and Technology of China
Academy of Engineering Physics under Grants Nos. 2010B0301048 and
2011A0301016.\
\end{acknowledgments}


\begin{thebibliography}{99}                                                                                               %


\bibitem {ref-1}S. Heathman, R. G. Haire, T. Le Bihan, A. Lindbaum, M. Idiri,
P. Normile, S. Li, R. Ahuja, B. Johansson, and G. H. Lander, Science
\textbf{309}, 110 (2005).

\bibitem {ref-2}I. D. Prodan, G. E. Scuseria, and R. L. Martin, Phys. Rev. B
\textbf{76}, 033101 (2007).

\bibitem {ref-3}R. Atta-Fynn and A. K. Ray, Phys. Rev. B \textbf{76}, 115101 (2007).

\bibitem {ref-04}K. T. Moore and G. van der Laan, Rev. Mod. Phys. \textbf{81},
235 (2009).

\bibitem {ref-5}J. M. Haschke, T.H. Allen, and L.A. Morales, Science
\textbf{287}, 285 (2000).

\bibitem {ref-6}M. T. Butterfield, T. Durakiewicz, E. Guziewicz, J. Joyce, A.
Arko, K. Graham, D. Moore, and L. Morales, Surf. Sci. \textbf{571}, 74 (2004).

\bibitem {ref-7}Gouder, A. Seibert, L. Havela, and J. Rebizant, Surf. Sci.
\textbf{601}, L77 (2007).

\bibitem {Has2000}J.M. Haschke, Los Alamos Science \textbf{26}, 253 (2000).

\bibitem {ref-4}M. T. Butterfield, T. Durakiewicz, I. D. Prodan, G. E.
Scuseria, E. Guziewicz, J. A. Sordo, K. N. Kudin, R. L. Martin, J. J. Joyce,
A. J. Arko, K. S. Graham, D. P. Moore, and L.A. Morales, Surf. Sci.
\textbf{600}, 1637 (2006).

\bibitem {ref-8}P. A. Korzhavyi, L.Vitos, D. A. Andersson and B. Johansson,
Nature Mater. \textbf{3}, 225 (2004).

\bibitem {ref-9}C. E. Boettger and A. K. Ray, Int. J. Quantum Chem.
\textbf{90}, 1470 (2002).

\bibitem {ref-10}C. McNeilly, J. Nucl. Mater. \textbf{11}, 53 (1964).

\bibitem {ref-11}S. L. Dudarev, D. N. Manh, and A. P. Sutton, Philos. Mag. B
\textbf{75}, 613 (1997).

\bibitem {ref-12}S. L. Dudarev, G. A. Botton, S. Y. Savrasov, C. J. Humphreys,
and A. P. Sutton, Phys. Rev. B \textbf{5}7, 1505 (1998).

\bibitem {ref-13}S. L. Dudarev, M. R. Castell, G. A. Botton, S. Y. Savrasov,
C. Muggelberg, G. A. D. Briggs, A. P. Sutton, and D. T. Goddard, Micron
\textbf{31}, 363 (2000).

\bibitem {ref-14}I. D. Prodan, G. E. Scuseria, J. A. Sordo, K. N. Kudin, and
R. L. Martin, J. Chem. Phys. \textbf{123}, 014703 (2005).

\bibitem {ref-15}L. Petit, A. Svane, Z. Szotek, W. M. Temmerman, and G. M.
Stocks, Phys. Rev. B \textbf{81}, 045108 (2010).

\bibitem {dmft}Q. Yin and S. Y. Savrasov, Phys. Rev. Lett. \textbf{100},
225504 (2008).

\bibitem {ref-16}B. Sun, P. Zhang, and X.-G. Zhao, J. Chem. Phys.
\textbf{128}, 084705 (2008).

\bibitem {ref-17}S. Bo and Z. Ping, Chin. Phys. B \textbf{17}, 1364 (2008).

\bibitem {ref-18}D. A. Andersson, J. Lezama, B. P. Uberuaga, C. Deo, and S. D.
Conradson, Phys. Rev. B \textbf{79}, 024110 (2009).

\bibitem {ref-19}G. Jomard, B. Amadon, F. Bottin, and M. Torrent, Phys. Rev. B
\textbf{78}, 075125 (2008).

\bibitem {ref-20}H. Shi, M. Chu, and P. Zhang, J. Nucl. Mater. \textbf{400},
151 (2010).

\bibitem {ref-21}P. Zhang, B.-T. Wang, and X.-G. Zhao, Phys. Rev. B
\textbf{82}, 144110 (2010).

\bibitem {SSR}M. V. G. Pirovano, A. Hofmann, and J. Sauer, Surf. Sci. Rep.
\textbf{62}, 219 (2007).

\bibitem {ref-22}X. Wu and A. K. Ray, Physica B \textbf{301}, 359 (2001).

\bibitem {ref-23}X. Wu and A. K. Ray, Phys. Rev. B \textbf{65}, 085403 (2002).

\bibitem {Vasp}G. Kresse and J. Furthm\"{u}ller, Phys. Rev. B \textbf{54},
11169 (1996).

\bibitem {Blo}P.E. Bl\"{o}chl, Phys. Rev. B \textbf{50}, 17953 (1994).

\bibitem {Paw}G. Kresse and D. Joubert, Phys. Rev. B \textbf{59}, 1758 (1999).

\bibitem {Perdew}J. P. Perdew, J. A. Chevary, S. H. Vosko, K. A. Jackson, M.
R. Pederson, D. J. Singh, and C. Fiolhais, Phys. Rev. B \textbf{46}, 6671 (1992).

\bibitem {Monk}H. J. Monkhorst and J. D. Pack, Phys. Rev. B \textbf{13}, 5188 (1976).

\bibitem {Tasker}P. W. Tasker, Solid State Phys. \textbf{12}, 4977 (1979).

\bibitem {Polar-1}H.-J. Freund, H. Kuhlenbeck, and V. Staemmler, Rep. Prog.
Phys. \textbf{59}, 283 (1996).

\bibitem {Polar-2}C. Noguera, J. Phys.: Condens. Matter \textbf{12}, R367 (2000).

\bibitem {Polar-3}C. Noguera and J. Goniakowski, J. Phys.: Condens. Matter
\textbf{20}, 264003 (2008).

\bibitem {Polar-4}C. Franchini, V. Bayer, R. Podloucky, G. Parteder, S.
Surnev, and F. P. Netzer, Phys. Rev. B \textbf{73}, 155402 (2006).

\bibitem {abAT-1}M. W. Finnis, Phys. Status Solidi A \textbf{166}, 397 (1998).

\bibitem {abAT-2}K. Reuter and M. Scheffler, Phys. Rev. B \textbf{65}, 035406 (2001).

\bibitem {chem-potential}\textit{NIST-JANAF Thermochemical Tables}, 4th ed.,
edited by J. Chase (American Chemical Society, Washington, DC, 1998).

\bibitem {GC-1}J. P. Perdew, A. Ruzsinszky, G. I. Csonka, O. A. Vydrov, G. E.
Scuseria, L. A. Constantin, X. Zhou, and K. Burke, Phys. Rev. Lett.
\textbf{100}, 136406 (2008).

\bibitem {GC-2}P. S\"{o}derlind and A. Gonis, Phys. Rev. B \textbf{82}, 033102 (2010).

\bibitem {UO2-zigzag}T. N. Taylor, W. P. Ellis, Surf. Sci. \textbf{107}, 249 (1981).

\bibitem {abAT-6}M. Fronzi, A. Soon, B. Delley, E. Traversa, and C. Stampfl,
J. Chem. Phys. \textbf{131}, 104701 (2009).

\bibitem {CeO2}N. V. Skorodumova, M. Baudin, and K. Hermansson, Phys. Rev. B
\textbf{69}, 075401 (2004).

\bibitem {UO2}H. Idriss, Surf. Sci. Rep. \textbf{65}, 67 (2010).

\bibitem {MgO}J. Jung, H.-J. Shin, Y. Kim, and M. Kawai, Phys. Rev. B
\textbf{82}, 085413 (2010).
\end{thebibliography}
\end{document}